\documentclass[preprint,12pt]{elsarticle}

\usepackage{graphicx}
\usepackage{amsmath}
\usepackage{amsfonts}
\usepackage{amssymb}
\usepackage{amsthm}
\usepackage{bbm}

\begin{document}

\begin{frontmatter}

\title{All-electrical manipulation of electron spin in a semiconductor nanotube}

\author[AGH]{P. W\'ojcik \corref{cor1}}
\ead{pawel.wojcik@fis.agh.edu.pl}
\cortext[cor1]{Corresponding author. Tel. +48 126174471}
\author[AGH]{J. Adamowski}
\author[AGH]{M. Wo{\l}oszyn}
\author[AGH]{B. J.  Spisak}

\address[AGH]{AGH University of Science and Technology,
Faculty of Physics and Applied Computer Science,
al.~Mickiewicza 30, 30-059~Krak\'ow, Poland}

\begin{abstract}

A controlled manipulation of electron spin states has been
investigated for a cylindrical two-dimensional electron gas
confined in a semiconductor nanotube/cylindrical nanowire with the
Rashba spin-orbit interaction. We present analytical solutions for
the two limiting cases, in which the spin-orbit interaction
results from (A) the radial electric field and (B) the electric
field applied along the $z$ axis of the nanotube.
In case (A), we have found that only the superposition of bands with the same
orbital momentum leads to the spin precession around
the cylinder (nanotube) axis.
In case (B), we have obtained the damped
oscillations of the $z$ spin component with the period that
changes as a function of the coordinate $z$. We have shown that
the damped oscillations of the average value of the $z$ spin
component form beats localized along the nanotube axis. 
The position of the beats can be controlled by the bias voltage.

\end{abstract}


\end{frontmatter}

\section{Introduction}
\label{sec:1}
The electron spin control induced by the electric field is a basic principle for the realization of
spintronic devices, including the spin transistor~\cite{Datta,Schliemann},
and the quantum operations on spin qubits~\cite{Szumniak,Berg}.
For this reason the spin-orbit interaction
(SOI)~\cite{Dresselhause,Rashba}, which couples the  momentum of the electron with its spin, has attracted the
substantial interest in recent years.
In the spintronic and quantum computing applications, the Rashba SOI~\cite{Rashba} is very promising,
since its strength can be controlled by the external electric field~\cite{Nitta,Matsuyama,Grundler},
which opens up a new possibility of manipulating the electron spin state by the gate voltages at zero
magnetic field~\cite{Nazmitdinov,Nowack,Meier}.
The spin manipulation due to the Rashba SOI has been recently demonstrated
in the electric-dipole spin resonance (EDSR) experiments performed in
the system of double quantum dots embedded in a gate InAs quantum wire~\cite{NadjPerge,NadjPerge2,Pfund}.
In the EDSR, the Pauli blockade of the current, which occurs when the quantum dots are occupied
by the electrons with parallel spins, is lifted by the Rashba SOI generated by the oscillating gate voltage.

Recently, the special attention has been directed towards non-planar low dimensional structures,
in which interesting physical effects resulting from the curvature of the surface have been found~\cite{Ferrari2}.
Many research studies are focused on the cylindrical two-dimensional electron gas (2DEG).
The nanostrucure containing the cylindrical 2DEG can be fabricated by self-rolling of a thin strained
semiconductor planar heterostructure grown by molecular-beam epitaxy~\cite{Vorobov, Mendach,Friedland}.
This method allows to obtain the free-standing semiconductor nanotubes with the radius that ranges
from several nanometers up to several micrometers.
The cylindrical 2DEG also appears in the core-shell nanowires, which are produced
on the cylindrical substrate with a multilayer overgrowth~\cite{Lauhon,Martensson}.
Another method of the fabrication of the cylindrical 2DEG
exploits the electrical neutrality which leads to the formation of the triangular quantum well
in the thin region near the surface of the semiconductor nanowire~\cite{Bjork}.
The electrons confined in this quantum well form the cylindrical 2DEG with radius of few nanometers.
The electronic properties of the cylindrical 2DEG are strongly dependent on the curvature,
which manifests themselves especially in the presence of the magnetic field.
The electron states have been recently calculated by Ferrari et al.~\cite{Ferrari} for the cylindrical 2DEG in the transverse
magnetic field. In this system, the electrons are coupled to the magnetic field component perpendicular to the surface that
varies along the circumference of the cylinder. The gradient of
magnetic field perpendicular to the surface causes that the electrons propagating  in the opposite directions become localized in
the opposite sides of the circumference~\cite{Kleiner}, which leads to the experimentally observed
Hall quantization~\cite{Friedland,Vorobov}.

The interplay between the curvature effects and the SOI in the
cylindrical 2DEG  has been studied in the recent papers~\cite{Wenk,Wohlman,Bringer}.
In Ref.~\cite{Wenk}, the authors investigated the dimensional
dependence of weak localization corrections and spin relaxation in
cylindrical nanowires with the Rashba SOI. The spin dependent
electric current through the cylindrical nanowire containing a
region with the spin-orbit coupling has been investigated by
Entin-Wohlman et al.~\cite{Wohlman} who have shown that the
tunneling through the region with the SOI causes that each step of
the quantized conductance splits into two separate steps with
the spin polarization perpendicular to the
direction of the current. The spin precession in the cylindrical
semiconductor nanowire due to the Rashba spin-orbit coupling has
been investigated by Bringer and Sch\"apers~\cite{Bringer}.
The authors~\cite{Bringer}
have taken into account the Rashba SOI generated by the radial electric field, which
results from the inhomogeneous radial redistribution of charge
near the surface of nanowire.

In the present paper,
we have included the Rashba SOI stemming from the axially directed electric field
that causes the flow of current and
studied the possibility of spin manipulation in the semiconductor nanotube
with the use of the radial electric field and the electric field acting along the
cylinder axis.
For both cases we have
obtained the analytical solutions and discussed them in the context of spin modulation.
The present results can be applied to both the semiconductor nanotubes with the few atomic monolayer thickness
and cylindrical nanowires with the 2DEG electron gas accumulated near the surface.

The paper is organized as follows: in Sec.~\ref{sec:2}, we describe the theoretical model of the cylindrical 2DEG
with the spin-orbit coupling, in Sec.~\ref{sec:3}, we present the results, the conclusions are presented in Sec~\ref{sec:4}.

\section{Theory}
\label{sec:2}

\begin{figure}[ht]
\begin{center}
\includegraphics[bb=23 360 830 830, scale=0.25,angle=0]{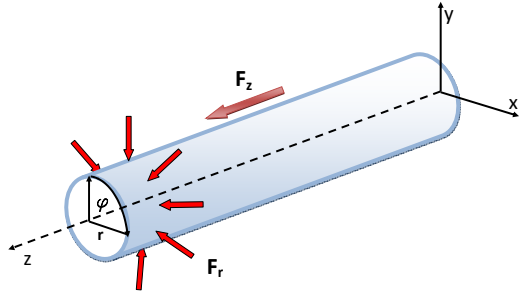}
\caption{Schematic of the nanotube with 2DEG.}
\label{fig1}
\end{center}
\end{figure}

We consider one-electron states in the cylindrical 2DEG
with the spin-orbit interaction that originates from the radial electric field $F_r$
and homogeneous electric field $F_z$ applied along the nanotube axis (Fig.~\ref{fig1}).

The Hamiltonian of the system takes on the form (see~\ref{append})
\begin{eqnarray}
\hat{H}=\left [ -\frac{\hbar ^2}{2m} \left ( \frac{1}{r_0^2} \frac{\partial ^2}{\partial \varphi ^2} +
\frac{\partial ^2}{\partial z ^2} \right ) -eF_z z \right ] \hat{\mathbb{I}}  + \hat{H}_{SO} \; ,
\label{RS}
\end{eqnarray}
where $\varphi$ and $z$ are the cylindrical coordinates (see Fig.~\ref{fig1}), $e$ is the elementary charge,
$r_0$ is the radius of the nanotube,
$m$ is the electron effective band mass,
$\hat{\mathbb{I}}$ is the $2\times 2$ unit matrix,
and $\hat{H}_{SO}$ is the Hamiltonian of the spin-orbit interaction.

The spin-orbit interaction couples the spin $\vec{s}$ of the electron
with its linear momentum $\vec{p}$ via the electric field $\vec{F}=-\nabla V/e$.
The SOI Hamiltonian can be expressed as (\ref{append})
\begin{equation}
\hat{H}_{SO}= -\frac{e\gamma}{\hbar} \vec{\sigma} \cdot (\vec{F} \times \vec{p})  \nonumber
 \end{equation}
\begin{equation}
=\left (
\begin{array}{cc}
-\frac{ie\alpha}{r_0}\frac{\partial}{\partial \varphi} &
e^{-i\varphi} \left ( -eF_z \frac{i\gamma}{r_0}\frac{\partial}{\partial \varphi}+e\alpha \frac{\partial}{\partial z} \right ) \\
e^{i\varphi} \left (-eF_z \frac{i\gamma}{r_0} \frac{\partial}{\partial \varphi} -  e\alpha \frac{\partial}{\partial z} \right ) &
\frac{ie\alpha}{r_0}\frac{\partial}{\partial \varphi}
\end{array}
\right )
 \end{equation}
where $\vec{\sigma}$ is the vector of Pauli matrices, $\vec{s}=(\hbar/2)\vec{\sigma}$ and
$\gamma$ is the coupling constant determined by the band structure
of the semiconductor.  In the present paper,
we will also use the effective coupling constant $\alpha=-\gamma F_r$ that takes
into account both the band structure  and the radial electric field effects.

The eigenstate of Hamiltonian (\ref{RS}) is the spinor with the two components corresponding
to the same total angular momentum $(l+1/2)(\hbar/2)$~\cite{Bringer}
\begin{equation}
 \Psi(\varphi,z)=
\left (
\begin{array}{c}
 \psi^{\uparrow}(\varphi,z) \\
\psi^{\downarrow}(\varphi,z)
\end{array}
\right ) =
\frac{e^{il\varphi}}{\sqrt{2\pi}}
\left (
\begin{array}{c}
f(z) \\
e^{i\varphi} g(z)
\end{array}
\right ) \;,
\label{spinor}
\end{equation}
where $l$ is the orbital quantum number.
After inserting (\ref{spinor}) into the eigenequation of Hamiltonian (\ref{RS})
we obtain
\begin{subequations}
\begin{eqnarray}
 &-&\frac{\hbar ^2}{2m} \frac{d^2 f(z)}{ dz^2}+\frac{\hbar ^2 l^2}{2mr_0^2} f(z)-eF_zzf(z)+\frac{e \alpha l}{r_0} f(z) \nonumber \\
&+&eF_z\frac{ \gamma(l+1)}{r_0}g(z)+e\alpha\frac{d g(z)}{d z}=Ef(z),
\label{RSIIa}
\end{eqnarray}
\begin{eqnarray}
 &-&\frac{\hbar ^2}{2m} \frac{d^2 g(z)}{ dz^2}+\frac{\hbar ^2 (l+1)^2}{2mr_0^2} g(z)-eF_zzg(z) \nonumber \\
 &-&\frac{e \alpha (l+1)}{r_0} g(z)+ eF_z\frac{ \gamma l}{r_0}f(z)-e\alpha\frac{d f(z)}{d z}=Eg(z),
\label{RSIIb}
\end{eqnarray}
\end{subequations}
where energy $E$ is measured with respect to the lowest energy of the size-quantized motion in the radial direction.\\

In general, the system of equations (\ref{RSIIa},~\ref{RSIIb}) is not solvable analytically.
Nevertheless, we have found that -- in the two limiting cases -- the analytical
solutions exist.  These are:
\begin{itemize}
 \item [(A)] Zero axial electric field ($F_z=0$), then the SOI is due to the radial electric field $F_r$,
 \item [(B)] Zero radial electric field ($F_r=0$, i.e., $\alpha=0$), then the SOI is due to the electric field $F_z$ applied along the nanotube axis.
\end{itemize}

\section{Results}
\label{sec:3}
In this section, we present the analytical solutions for
the InAs nanotube/cylindrical nanowire, for which we take on the following values of
the parameters:  electron band effective mass  $m=0.026m_0$, where $m_0$ is the free electron mass,
the  radius of the nanotube $r_0=50$~nm, and the spin-orbit interaction constants
$\gamma=1.17$~nm$^2$ and $\alpha=10$~meVnm~\cite{Roulleau,Winkler}.
We discuss the results obtained for two limiting cases (A) and (B).
In both the cases, the analytical solutions are
applied to describe a possible control of spin precession.

\subsection{Effect of radial electric field}

If the axial electric field is equal to zero ($F_z=0$)
and the electric field has only the radial component, the solution
of Eqs.~(\ref{RSIIa}) and (\ref{RSIIb}) takes on the form
\begin{equation}
\left (
\begin{array}{c}
f(z) \\
g(z)
\end{array}
\right ) =
\left (
\begin{array}{c}
u_l \\
v_l
\end{array}
\right )
e^{ikz} \; ,
\label{eq:c1r1}
\end{equation}
where $k \equiv k_z$ is the $z$ component of the wave vector and $u_l$ and $v_l$ are the amplitudes of the spin-up and
spin-down spinor elements for the subband with orbital quantum number $l$.\\
If we introduce energy $\varepsilon$ of the electron with the subtracted kinetic energy
related to its motion along the $z$ axis, i.e.,
$\varepsilon=E-\hbar ^2 k^2/(2m)$, where $E$ is the total energy, the system of equations
(\ref{RSIIa}-\ref{RSIIb}) reduces  to
\begin{subequations}
\begin{equation}
V_l u_l+\frac{e \alpha l}{r_0}u_l+iek\alpha v_l= \varepsilon u_l \;,
\label{sys2} \\
\end{equation}
\begin{equation}
V_{l+1} v_l-\frac{e \alpha (l+1)}{r_0}v_l-iek\alpha u_l= \varepsilon v_l \;,
\label{sys1}
\end{equation}
\end{subequations}
where $V_l=\hbar ^2 l^2/(2mr_0^2)$. \\
The system of equations (\ref{sys2}-\ref{sys1}) has non-trivial solutions
only if the following characteristic equation is satisfied
\begin{equation}
 \varepsilon ^2+ \left[ \frac{e \alpha}{r_0} - \left( V_l+V_{l+1} \right) \right ] \varepsilon
+ \omega = 0\: ,
\label{eq:c1r2}
\end{equation}
where
\begin{equation}
\omega= V_lV_{l+1}-V_l\frac{e \alpha(l+1)}{r_0}
+V_{l+1}\frac{e \alpha l}{r_0}-\frac{ e^2 \alpha ^2 l(l+1)}{r_0^2}-e^2k^2\alpha^2.
\end{equation}
The quadratic equation~(\ref{eq:c1r2}) possesses two solutions $\varepsilon ^{\pm}$ that can be
obtained by the simple analytical calculation.
The difference between them $\Delta \varepsilon=\varepsilon ^+ - \varepsilon ^-$
contains the spin-orbit energy splitting and the angular momentum contribution,
and is given by
\begin{equation}
\Delta \varepsilon= \frac{1}{2}\sqrt{\left[ \frac{e \alpha}{r_0} - \left( V_l+V_{l+1} \right) \right ]^2-4\omega}.
\end{equation}
Using the normalization condition $|u_l|^2+|v_l|^2=1$ we obtain the
amplitudes $u_l^{\pm}$ and  $v_l^{\pm}$ that correspond to the energies $\varepsilon ^{\pm}$
\begin{eqnarray}
\label{eqcof1}
u^{\pm}_l&=&-\frac{iek\alpha}{V_l+\frac{e\alpha l}{r_0}-\varepsilon ^{\pm}}v_l^{\pm}\;, \\
|v_l^{\pm}|^2&=&\frac{|V_l+\frac{e\alpha l}{r_0}-\varepsilon ^{\pm}|^2}{|V_l+\frac{e\alpha l}{r_0}-\varepsilon ^{\pm}|^2+|ek\alpha|^2} \;.
 \label{eqcof2}
\end{eqnarray}

\begin{figure}[ht]
\begin{center}
\includegraphics[scale=.2]{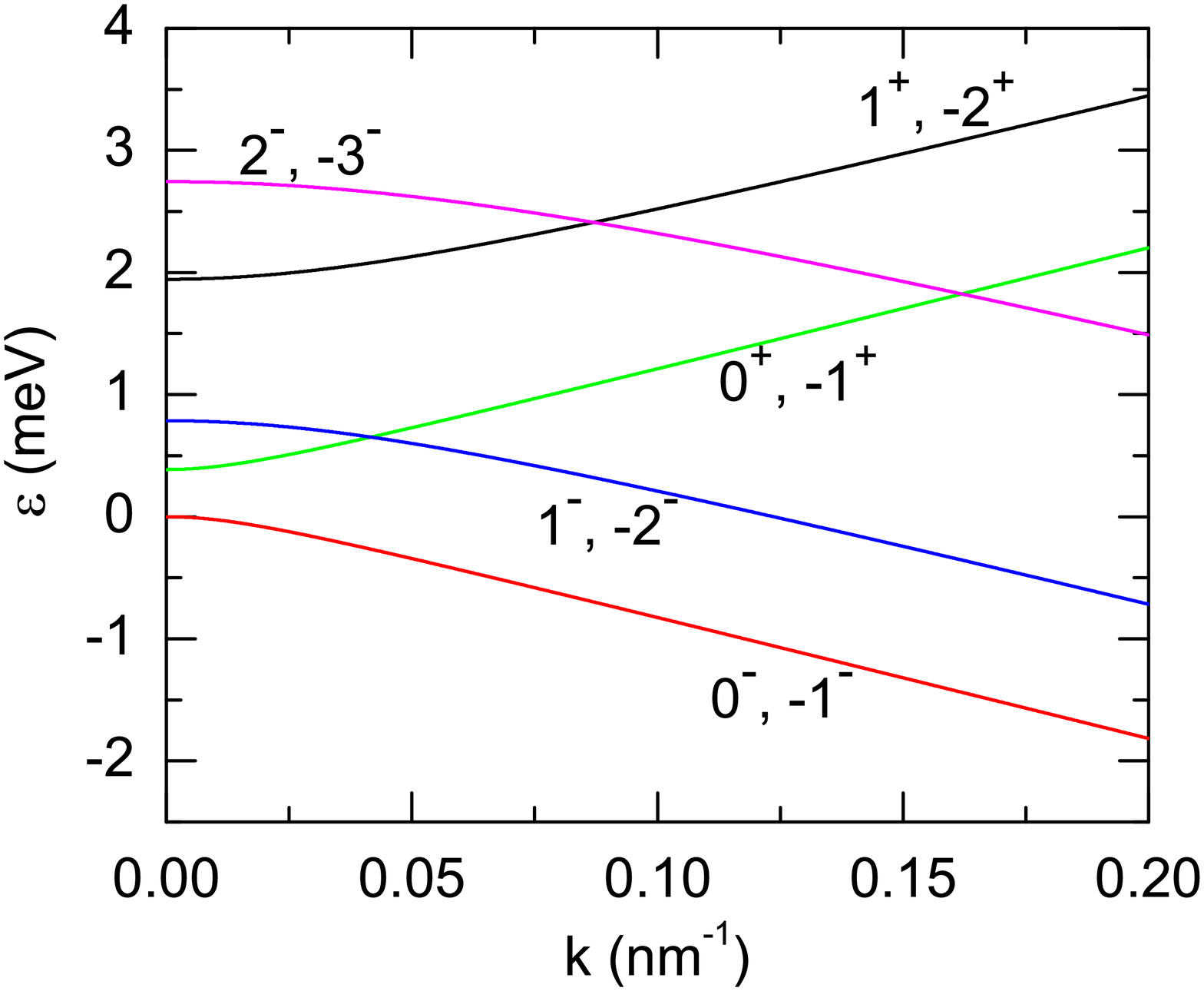}
\caption{Energy $\varepsilon$ as a function of wave vector $k$ and
orbital quantum number $l$.
The energy band with orbital momentum $\hbar l$ and energy $\varepsilon^{\pm}$ is denoted by $l^{\pm}$.}
\label{fig2}
\end{center}
\end{figure}

\begin{figure}[ht]
\begin{center}
\includegraphics[scale=.4]{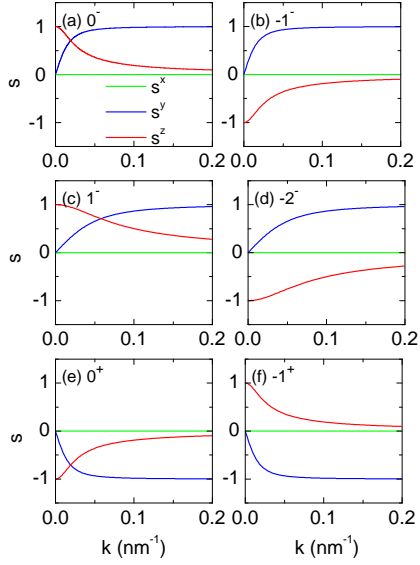}
\caption{Expectation values of spin components (in units $\hbar/2$)  as  functions of wave vector $k$
for several subbands $l^{\pm}$.}
\label{fig3}
\end{center}
\end{figure}

In Fig.~\ref{fig2}, we have plotted the energy $\varepsilon^{\pm}$ as a function of the wave vector $k$ for several
orbital quantum numbers $l$.
The energy band with orbital momentum $\hbar l$ and energy $\varepsilon^{\pm}$ is denoted by $l^{\pm}$.
We see that for given orbital quantum number $l$ the energy $\varepsilon ^+$ increases with increasing wave vector $k$,  while
the energy $\varepsilon ^-$ decreases as a function of $k$.
In the following, the wave functions corresponding to orbital quantum number $l$
and energy $\varepsilon^{\pm}$ are denoted by $\psi_{l}^{\pm}(\varphi,z)$.
Since the spin-orbit interaction does not lift the
time reversal symmetry, the Kramers degeneracy of states $\psi_l^{\pm}(\varphi,z)$ and $\psi_{-(l+1)}^{\pm}(\varphi,z)$
is clearly visible  in Fig.~\ref{fig2}.
The expectation values of the spin components
$s_{l^{\pm}}^{x,y,z}= \langle \psi_l^{\pm}(\varphi,z) |\hat{s}^{x,y,z}|\psi_l^{\pm}(\varphi,z) \rangle_{\varphi}$,
where $\langle \ldots \rangle_{\varphi}$ denotes the integration over the angle $\varphi$,
are given by the following analytical formulas
\begin{subequations}
\begin{equation}
 s_{l^{\pm}}^x= \frac{\hbar}{2} \left ( u^{*\pm}_{l}v^{\pm}_l+v^{*\pm}_{l}u^{\pm}_l \right ),
 \label{eqsx}
\end{equation}
 \begin{equation}
 s_{l^{\pm}}^y=\frac{\hbar}{2} \left ( -iu^{*\pm}_{l}v^{\pm}_l+iv^{*\pm}_{l}u^{\pm}_l \right ),
 \end{equation}
\begin{equation}
 s_{l^{\pm}}^z=\frac{\hbar}{2} \left ( u^{*\pm}_{l}u^{\pm}_l-v^{*\pm}_{l}v^{\pm}_l \right ) \;.
 \end{equation}
\end{subequations}
In Fig.~\ref{fig3}, the expectation values of the spin components are displayed
as functions of the wave vector $k$.

We see that  $s^x_{l^{\pm}}$ is equal to zero and does not depend on the wave
vector $k$ for all subbands. This features results directly from eqs.~(\ref{eqcof1}), (\ref{eqcof2}), and (\ref{eqsx}).
The different behavior is observed for  the $s^y_{l^{\pm}}$ and $s^z_{l^{\pm}}$ spin components,
which change as a function of wave vector $k$, whereas
$s^z_{l^{\pm}}$  decreases and $s^y_{l^{\pm}}$ increases with $k$.
Fig.~\ref{fig3} shows that the states with the same angular momenta $l$ and different energy $\varepsilon ^{\pm}$
possess spin components $s^y_{l^{\pm}}$ and $s^z_{l^{\pm}}$  with opposite signs [cf. Figures~\ref{fig3}(a) and (e)].

Through the present paper we focus on the spin precession induced by the Rashba spin-orbit interaction.
For this purpose we consider the quantum state, which is the superposition of one-electron states $\psi_{l}^{\pm}(\varphi,z)$.
The superposition of the states with orbital quantum numbers $l_1$ and $l_2$  can be expressed
using the Bloch sphere representation as follows:
\begin{equation}
\Psi_{l_1^{\pm},l_2^{\pm}}(\varphi,z)=\cos \left ( \frac{\theta}{2} \right ) \psi_{l_1}^{\pm}(\varphi,z)
+\sin \left ( \frac{\theta}{2} \right ) e^{i\phi} \psi_{l_2}^{\pm}(\varphi,z) \; ,
\label{eq:BS}
\end{equation}
where the angles $\theta$ and $\phi$ determine the quantum state on the Bloch sphere.
For the eigenstates given by eq.~(\ref{eq:BS}), the expectation values of the spin components
$s^{x,y,z}_{l_1^{\pm},l_2^{\pm}}=\langle \Psi_{l_1^{\pm},l_2^{\pm}}(\varphi,z) |\hat{s}^{x,y,z}| \Psi_{l_1^{\pm},l_2^{\pm}}(\varphi,z) \rangle _{\varphi}$
are expressed as follows:
\begin{subequations}
 \begin{eqnarray}
  &&s^x_{l_1^{\pm},l_2^{\pm}}=\cos ^2 \left ( \frac{\theta}{2} \right ) s^x_{l_1^{\pm}} + \sin ^2 \left ( \frac{\theta}{2} \right ) s^x_{l_2^{\pm}} +\frac{\hbar}{2} \sin \theta \nonumber \\
  && \times \Re \left [ e^{i \phi}  (u^{*\pm}_{l_1}v^{\pm}_{l_2}+v^{*\pm}_{l_1}u^{\pm}_{l_2})
  e^{-i(k_{l_1}^{\pm}-k_{l_2}^{\pm})z}  \delta_{l_1,l_2}  \right ] \;,
  \label{eq:sssx}
 \end{eqnarray}
 \begin{eqnarray}
  &&s^y_{l_1^{\pm},l_2^{\pm}}=\cos ^2 \left (\frac{\theta}{2} \right ) s^y_{l_1^{\pm}} + \sin ^2 \left ( \frac{\theta}{2} \right ) s^y_{l_2'^{\pm}} +\frac{\hbar}{2} \sin \theta \nonumber \\
  && \times \Re \left [ e^{i \phi}  (-iu^{*\pm}_{l_1}v^{\pm}_{l_2}+iv^{*\pm}_{l_1}u^{\pm}_{l_2})
  e^{-i(k_{l_1}^{\pm}-k_{l_2}^{\pm})z} \delta_{l_1,l_2}  \right ] \;,
 \label{eq:sssy}
 \end{eqnarray}
  \begin{eqnarray}
  &&s^z_{l_1^{\pm},l_2^{\pm}}=\cos ^2 \left ( \frac{\theta}{2} \right ) s^z_{l_1^{\pm}} + \sin ^2 \left ( \frac{\theta}{2} \right ) s^z_{l_2^{\pm}} +\frac{\hbar}{2} \sin \theta \nonumber \\
  &&  \times \Re \left [ e^{i \phi}  (u^{*\pm}_{l_1}u^{\pm}_{l_2}-v^{*\pm}_{l_1}u^{\pm}_{l_2})
  e^{-i(k_{l_1}^{\pm}-k_{l_2}^{\pm})z} \delta_{l_1,l_2} \right ],
  \label{eq:sssz}
 \end{eqnarray}
\end{subequations}
where $\Re$ denotes the real part of the complex number and
$\delta_{l_1,l_2}= \left ( \int _{0}^{2\pi} e^{-i(l_1-l_2)\varphi}\right) / 2\pi$ is the Kronecker delta.

In formulas (\ref{eq:sssx}-\ref{eq:sssz}), $k_{l_1}^{\pm}$ and $k_{l_2}^{\pm}$ are the wave vectors corresponding to the
one-electron states $\psi_{l_1}^{\pm}(\varphi,z)$ and $\psi_{l_2}^{\pm}(\varphi,z)$,
which are the components of the superposition state $\Psi_{l_1^{\pm},l_2^{\pm}}(\varphi,z)$.
For the fixed total electron energy $E$, the wave vectors $k_{l_1}^{\pm}$ and $k_{l_2}^{\pm}$ are different.
The non-vanishing difference $\Delta k=|k_{l_1}^{\pm}-k_{l_2}^{\pm}|$ causes that the orientation of the average spin vector
in the superposition state changes continuously along the $z$ axis [see eqs.~(\ref{eq:sssx}-\ref{eq:sssz})].
Throughout the present paper, we call such changes of the spin orientation the spin precession.
The precession length is determined as follows: $\lambda_{SO}=2\pi / \Delta\! k$.
The analysis of eqs.~(\ref{eq:sssx}-\ref{eq:sssz}) allows us to conclude that the spin precession along the cylinder axis
appears in the superposition states $\Psi_{{l}^{\pm},{l}^{\pm}}(\varphi,z)$, which are the combinations of the states with the same angular momenta.
This conclusion results directly from the analytical expressions (\ref{eq:sssx}-\ref{eq:sssz}) and supplements
the numerical results reported by Bringer and Sch\"apers~\cite{Bringer}, according to which the spin precession
around the cylinder axis appears in  the superposition of the two states with different total angular momenta.
\begin{figure}[ht]
\begin{center}
\includegraphics[scale=.3]{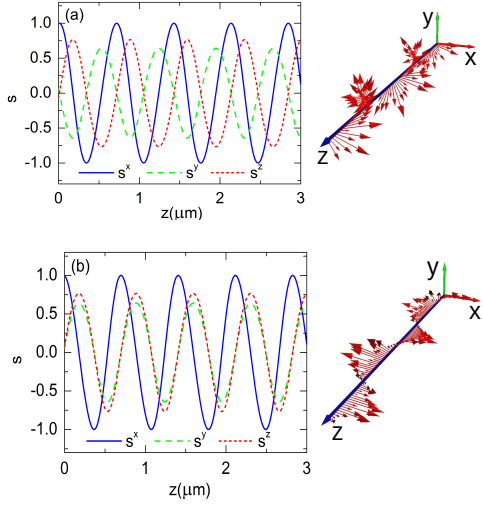}
\caption{Precession of spin components $s^{x,y,z}$ (in units $\hbar/2$) calculated for the
electron energy $E=1$~meV.
The results of calculations for the state (a)~$\Psi_{0^-,0^+}(\varphi,z)$  and (b)~$\Psi_{-1^--1^+}(\varphi,z)$.
}
\label{fig4}
\end{center}
\end{figure}

Let us consider now the three lowest-energy bands (see Fig.~\ref{fig2}).
According to the above finding only the superposition states
$\Psi_{0^-,0^+}(\varphi,z)$ and $\Psi_{-1^-,-1^+}(\varphi,z)$ contribute to the spin precession.
We note that both these states are degenerate, which results from the time reversal symmetry (Kramers degeneracy).
Fig.~\ref{fig4} displays the precession of the spin components $s^{x,y,z}$
for both the considered states $\Psi_{0^-,0^+}(\varphi,z)$ and $\Psi_{-1^-,1^+}(\varphi,z)$.
In the present calculations, the values of parameters $\theta$ and $\phi$ have been chosen so that for $z=0$ the spin
of the injected electron is parallel to the $x$ axis.
Fig.~\ref{fig4} shows that in the both states the electron spin precesses
around the cylinder axis, but the directions of the precession in the $x-y$ plane are opposite.
Namely, the spin rotates clockwise in state $\Psi_{0^-,0^+}(\varphi,z)$ and counterclockwise in state
$\Psi_{-1^-,-1^+}(\varphi,z)$. Moreover, the spin precession in the $x-y$ plane is accompanied by the
rotation of the $s^z$ spin component.
\begin{figure}[ht]
\begin{center}
\includegraphics[scale=.2]{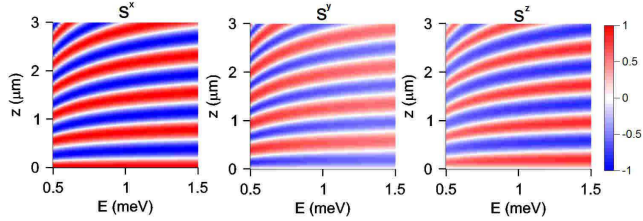}
 \caption{Precession of spin components $s^{x,y,z}$ (in units $\hbar/2$) as a function of energy calculated for state
$\Psi_{0^-,0^+}(\varphi,z)$ }
\label{fig5}
\end{center}
\end{figure}
Since $\Delta k$ changes with the electron energy, the precession length $\lambda_{SO}$
depends on the energy of the injected electrons. In Fig.~\ref{fig5}, we present the precession of the spin components
as a function of energy for the superposition state $\Psi_{0^-,0^+}(\varphi,z)$. We see that the precession length $\lambda_{SO}$
decreases with the increasing electron energy.
\begin{figure}[ht]
\begin{center}
\includegraphics[scale=.35]{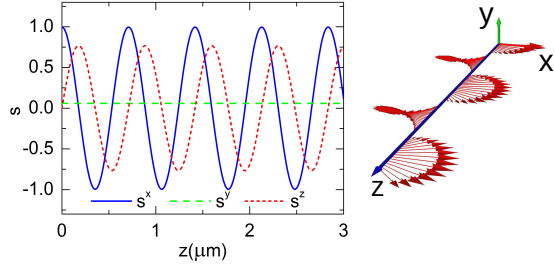}
\caption{Precession of electron spin in state $\Psi(\varphi,z)$.}
 \label{fig6}
\end{center}
\end{figure}

Since the electron with  total energy $E$ can not be injected only in one of the degenerate states,
we consider the precession of electron spin in the state, which is the superposition of states $\Psi_{0^-,0^+}(\varphi,z)$
and $\Psi_{-1^-,-1^+}(\varphi,z)$, i.e.,
$\Psi(\varphi,z)= \left ( \Psi_{0^-,0^+}(\varphi,z)+\Psi_{-1^-,-1^+}(\varphi,z) \right )/\sqrt{2}$.
In Fig.~\ref{fig6}, we present the changes of electron spin in state $\Psi(\varphi,z)$ along the cylinder axis.
We see that the precession of the electron spin in the $x-y$ plane vanishes, which results from the opposite directions of the precession
in states  $\Psi_{0^-,0^+}(\varphi,z)$  and $\Psi_{-1^-,-1^+}(\varphi,z)$ (cf. Fig.~\ref{fig4}).
Nevertheless, we still obtain the changes of $s^x$ and $s^z$ spin components. This fact is of a crucial importance for the electrical
control of electron spin in the nanotube and means that the electron spin can be efficiently
modulated by the radial electric field.

\subsection{Effect of axial electric field}
In this subsection, we study the influence of the longitudinal electric field on the spin precession
in the nanotube/nanowire.
We consider the nanotube/nanowire, in which the homogeneous electric field $F_z$ is applied parallel to the $z$ axis.
If we assume that total energy $E$ is a sum of kinetic energy $E_k$
of the longitudinal motion and energy $\varepsilon$ of the spin-orbit interaction, i.e., $E=E_k+\varepsilon$,
the solution of equation system (\ref{RSIIa}) and (\ref{RSIIb}) takes on the form
\begin{equation}
\left (
\begin{array}{c}
f(z) \\
g(z)
\end{array}
\right ) =
\left (
\begin{array}{c}
u_l \\
v_l
\end{array}
\right )
\mathcal{A}(-\xi) \;,
\label{eq:s2s}
\end{equation}
where $\mathcal{A}(\xi)$ is the Airy function and
\begin{equation}
\xi=\left (\frac{2meF_z}{\hbar ^2} \right )^{1/3} \left ( z+ \frac{E_k}{eF_z} \right ). \;
\label{xi}
\end{equation}
In this case, the system of equations (\ref{RSIIa}-\ref{RSIIb}) reduces to
\begin{subequations}
\begin{equation}
V_l u_l+e\gamma \frac{l+1}{r_0}eF_z v_l= \varepsilon  u_l,
\label{seq1}
\end{equation}
\begin{equation}
V_{l+1} v_l+e\gamma \frac{l}{r_0}eF_z u_l= \varepsilon  v_l.
\label{seq2}
\end{equation}
\end{subequations}
Equations (\ref{seq1}-\ref{seq2})
possess the non-trivial solutions only for the following two different values of energy
\begin{eqnarray}
\varepsilon^{\pm}&=&\frac{1}{2} \bigg [ (V_l+V_{l+1}) \nonumber \\
&& \pm \sqrt{(V_l-V_{l+1})^2+4 \gamma ^2 e^4 F_z^2 \frac{l(l+1)}{r_0^2} } \bigg ] \;.
\label{eq:eF}
\end{eqnarray}
Using the normalization condition $|u_l|^2+|v_l|^2=1$ we derive the analytical
expressions for coefficients $u_l^{\pm}$ and $v_l^{\pm}$ that correspond to the energy values $\varepsilon ^{\pm}$,
respectively,
\begin{eqnarray}
u_l^{\pm}&=&-\frac{\beta}{V_l-\varepsilon ^{\pm}}v_0^{\pm} \;, \\
 |v_l^{\pm}|^2&=&\frac{(V_l-\varepsilon ^{\pm})^2}{(V_l-\varepsilon ^{\pm})^2+ \beta^2} \;,
\label{seq3}
\end{eqnarray}
where $\beta = \gamma e^2 F_z (l+1)/r_0$.

\begin{figure}[ht]
\begin{center}
\includegraphics[scale=0.4]{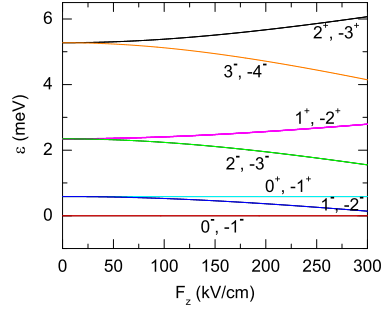}
\caption{ Energy $\varepsilon$ as a function of electric field $F_z$.
The results for the states with orbital quantum number $l$ and energy $\varepsilon ^{\pm}$
are marked by $l^{\pm}$.}
\label{fig7}
\end{center}
\end{figure}

In Fig.~\ref{fig7}, we display
energy $\varepsilon$ as a function of electric field $F_z$ for several $l$-bands.
We see that for the electric field $F_z=0$ each state is fourfold
degenerate, which results from the following symmetries: the
$z$-parity symmetry ($z \leftrightarrow -z$) leading to the
degeneracy of states $\psi_l^+(\varphi,z)$ and
$\psi_{l+1}^-(\varphi,z)$ and the time reversal symmetry leading
to the degeneracy of states with angular quantum number
$l$ and $-(l+1)$ with the opposite spins (Kramers degeneracy).
If the electric field is switched on, the $z$-parity symmetry is broken
and the degeneracy of states $\psi_l^{+}(\varphi,z)$ and $\psi_{l+1}^{-}(\varphi,z)$
is lifted. However, the electric field does
not break the time reversal symmetry, which means that the states
$\psi_{l}^\pm(\varphi,z)$ and $\psi_{-(l+1)}^\pm(\varphi,z)$ remain degenerate.
In order to study the influence of the homogeneous electric field applied along the cylinder axis
on the spin of the electron, we consider the superposition of states
given by eq.~(\ref{eq:BS}). The expectation values of spin components in  state $\Psi_{l_1^{\pm},l_2^{\pm}}(\varphi,z)$
are given by
\begin{subequations}
 \begin{eqnarray}
  &&s^x_{l_1^{\pm},l_2^{\pm}}=\cos ^2 \left ( \frac{\theta}{2} \right ) s^x_{l_1^{\pm}}
  + \sin ^2 \left ( \frac{\theta}{2} \right ) s^x_{l_2^{\pm}}+ \frac{\hbar}{2} \sin \theta \nonumber \\
 && \times \Re \left [ e^{i \phi}  (u^{*\pm}_{l_1}v^{\pm}_{l_2}+v^{*\pm}_{l_1}u^{\pm}_{l_2})
 \mathcal{A}(-\xi_{l_1}^{\pm})\mathcal{A}(-\xi_{l_2}^{\pm}) \delta_{l_1,l_2}  \right ] \;,
  \label{eq:sssx2}
 \end{eqnarray}
 \begin{eqnarray}
  &&s^y_{l_1^{\pm},l_2^{\pm}}=\cos ^2 \left (\frac{\theta}{2} \right ) s^y_{l_1^{\pm}}
  + \sin ^2 \left ( \frac{\theta}{2} \right ) s^y_{l_2^{\pm}}+ \frac{\hbar}{2} \sin \theta  \nonumber \\
  && \times \Re \left [ e^{i \phi}  (-iu^{*\pm}_{l_1}v^{\pm}_{l_2}+iv^{*\pm}_{l_1}u^{\pm}_{l_2})
  \mathcal{A}(-\xi_{l_1}^{\pm})\mathcal{A}(-\xi_{l_2}^{\pm}) \delta_{l_1,l_2}  \right ] \;,
 \label{eq:sssy2}
 \end{eqnarray}
  \begin{eqnarray}
  &&s^z_{l_1^{\pm},l_2^{\pm}}=\cos ^2 \left ( \frac{\theta}{2} \right ) s^z_{l_1^{\pm}}
  + \sin ^2 \left ( \frac{\theta}{2} \right ) s^z_{l_2^{\pm}}+ \frac{\hbar}{2} \sin \theta  \nonumber \\
  &&  \times \Re \left [ e^{i \phi}  (u^{*\pm}_{l_1}u^{\pm}_{l_2}-v^{*\pm}_{l_1}u^{\pm}_{l_2})
  \mathcal{A}(-\xi_{l_1}^{\pm})\mathcal{A}(-\xi_{l_2}^{\pm}) \delta_{l_1,l_2} \right ] \;.
  \label{eq:sssz2}
 \end{eqnarray}
\end{subequations}
In the formulas (\ref{eq:sssx2}-\ref{eq:sssz2}), $\xi_{l_1}^{\pm}$ and $\xi_{l_2}^{\pm}$ are the arguments of the Airy function for the
single electron states $\psi_{l_1}^{\pm}(\varphi,z)$ and $\psi_{l_2}^{\pm}(\varphi,z)$, which are the components of the superposition states.
We see that in the case of the longitudinal  electric field,  the spin precession about the
cylinder axis can be also observed only in the superposition states $\Psi_{l^{\pm},l^{\pm}}(\varphi,z)$, which are
the linear combination of states with the same angular momenta.
\begin{figure}[ht]
\begin{center}
\includegraphics[scale=0.5]{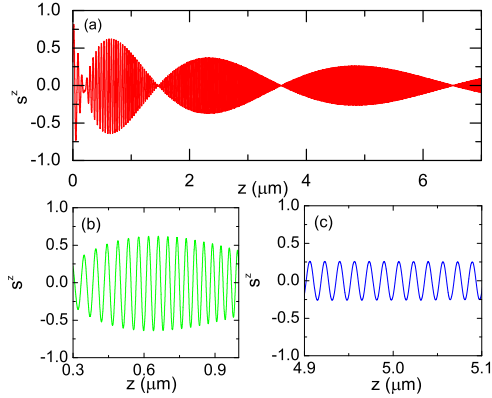}
\caption{(a) Average value $s^z$ of $z$-spin component in superposition state
$\Psi_{0^+,0^-}(\varphi,z)$  calculated for the electric
field $F_z=0.1$~kV/cm. Panels (b) and (c) are the zooms of panel (a) for the different position ranges.
Spin is measured in units $\hbar/2$.}
\label{fig8}
\end{center}
\end{figure}
Based on the results presented in sec.~3.1, which show that the superposition of both the Kramers degenerate states $\Psi_{0^+,0^-}(\varphi,z)$
and $\Psi_{-1^+,-1^-}(\varphi,z)$ leads to the changes of $s^x$ and $s^z$ components,  we
restrict our analysis of the spin precession in one of the degenerate state $\Psi_{0^+,0^-}(\varphi,z)$.
Fig.~\ref{fig8}(a) displays the average value of the $z$ spin component as a function of the position in the nanotube
calculated for the state  $\Psi_{0^+,0^-}(\varphi,z)$.
\begin{figure}[ht]
\begin{center}
\includegraphics[scale=.2]{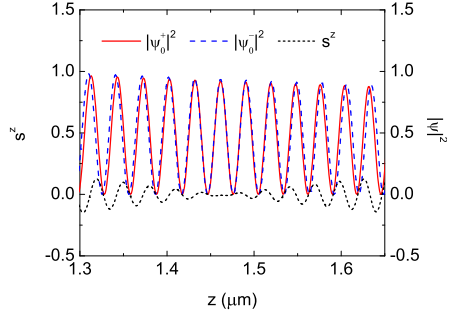}
\caption{Probability density $|\psi_{0}^{+}(\varphi,z)|^2$ and $|\psi_{0}^{-}(\varphi,z)|^2$ (solid and dashed curves)
and the average value $ \langle s_z \rangle$ of $z$ spin component (dotted curve) displayed in the vicinity of the position
($z \simeq 1.5 \mu$m), at which the average value of the $z$ spin component disappears. Spin is measured in units $\hbar/2$.}
\label{fig9}
\end{center}
\end{figure}
If the total energy $E$ of the electron is fixed, the difference between energies $\varepsilon^{+}$ and $\varepsilon^{-}$
causes that the kinetic energies $E_k$ corresponding to the states $\psi_{0}^{+}(\varphi,z)$ and $\psi_{0}^{-}(\varphi,z)$, which
form the superposition state $\Psi_{0^+,0^-}(\varphi,z)$, are different.
Since the kinetic energy appears in the argument of the Airy function
in solution (\ref{eq:s2s}), the kinetic energy difference
causes that the corresponding states are shifted in phase. This shift leads to the oscillatory changes of the average
$z$ spin component in the superposition state $\Psi_{0^+,0^-}(\varphi,z)$ (cf. Fig.~\ref{fig8}).
Fig.~\ref{fig8} demonstrates the periodic features that are closely related to the properties of the Airy function and the fact that
for $z\rightarrow - \infty$ potential energy $-eFz \rightarrow \infty$, which means that
the considered system is open only on one side.
The open boundary conditions  lead to the monotonic decrease of the average value of the $z$ spin component
with coordinate $z$, which results from the monotonically decreasing Airy function $\mathcal{A}(\xi)$.

The appearance of beats [Fig.~\ref{fig8}(a)] is a very interesting property of the damped oscillations of the average $z$-spin component.
Fig.~\ref{fig8}(a) shows that in some points on the nanowire axis $ s^z$
 vanishes and in the other points exhibits the local maxima.
In order to get a more deep physical insight into these phenomena,  in Fig.~\ref{fig9} we present
the probability density calculated for states $\psi_{0}^{+}(\varphi,z)$ and $\psi_{0}^{-}(\varphi,z)$, averaged over
coordinate $\varphi$, together with the average value of the $z$-spin component.
We see that the average $z$ spin component vanishes in the positions, in which
electron probability densities $|\psi_{0}^{+}(z)|^2$ and $|\psi_{0}^{-}(z)|^2$ oscillate in phase.
The zeroing of  $ s^z $
results from the fact that in states
$\psi_{0}^{+}(\varphi,z)$ and $\psi_{0}^{-}(\varphi,z)$, which are the components of $\Psi_{0^+,0^-}(\varphi,z)$,
the electrons possess the opposite spins.  Therefore, the integration over $\varphi$ in the right-hand side of the formula
\begin{equation}
 s_z = \frac{\hbar}{2} \langle \Psi_{0^+,0^-}(\varphi, z) | \sigma_z | \Psi_{0^+,0^-}(\varphi, z) \rangle _{\varphi}
\end{equation}
leads to the expression
\begin{equation}
 s_z  = \frac{\hbar}{2} \left ( |\psi_{0}^{+}(z)|^2-|\psi_{0}^{-}(z)|^2 \right )
\end{equation}
that is equal zero for these positions $z$, for  which electron probability densities $|\psi_{0}^{+}(z)|^2$ and $|\psi_{0}^{-}(z)|^2$
are equal to each other.\\
Since the oscillation period of  $|\psi_{0}^{+}(z)|^2$ and $|\psi_{0}^{-}(z)|^2$ increases with increasing coordinate $z$,
(which results from the property of the Airy function)
the positions, at which the average value of the $z$-spin component vanishes, are not equidistant.
\begin{figure}[ht]
\begin{center}
\includegraphics[scale=0.4]{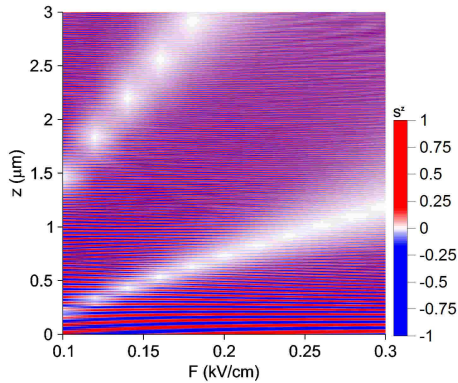}
\caption{Average value $s^z$ of $z$ spin component (in units $\hbar/2$) in the superposition state  $\Psi_{0^+,0^-}(\varphi,z)$
as a function of electric field $F_z$.}
\label{fig10}
\end{center}
\end{figure}
For the same reason the spin oscillation period is not constant but changes along the $z$ axis [cf. Figs.\ref{fig8}(b) and \ref{fig8}(c)].
The phase shift between both the components of the wave function depends on the spin-orbit interaction.
Therefore, the positions of the beats change if the electron energy or/and the electric field are changed.
In Fig.~\ref{fig10}, the average value of $z$ spin component is plotted as a function of axial
electric field $F_z$. The white bands correspond to the positions of the beats.
We see that the zeros and maxima (minima) of the average spin component can be shifted along the nanowire axis
if we change the external electric field.  This property opens up a possibility to control of the electron spin by
tuning the bias voltage.

\section{Conclusions}
\label{sec:4}

In the cylindrical nanowire, the 2DEG can be realized by confining the electrons in a narrow
asymmetric quantum well created near the surface of nanowire.
The reduction of the radial degree of freedom leads to the quantization of
the radial electron motion.
If the  potential confining the electrons near the surface is sufficiently strong,
all the electrons occupy the lowest-energy state, which results in the creation of the cylindrical 2DEG.
The asymmetry of this quantum well causes that the cylindrical 2DEG is subjected to the electric field,
which acts perpendicularly to the surface of the cylinder.
In order to generate the  flow of electrons through the nanowire, the electric field
parallel to the nanowire axis has to be applied.
Both the electric fields couple the momentum of the electron with its spin via the Rashba SOI
and should be taken into account when describing the electron spin effects in the cylindrical nanotubes and nanowires.

In this paper, we present the analytical solutions for two limiting cases for which
the spin-orbit interaction is generated by (A) the radial electric field
and (B) the electric field applied along the $z$ axis.
If the spin-orbit interaction is only due to the  radial electric field (case A),
the superposition of the electron states with the same angular momenta leads to the precession of the expectation
value of the electron spin in the $x-y$ plane. The precession in the $x-y$ plane vanishes
if the electron is injected in the superposition state,
which is the linear combination of both the degenerate states $\Psi_{0^-,0^+}(\varphi,z)$ and $\Psi_{-1^-,-1^+}(\varphi,z)$.
Nevertheless we still obtain the changes of $s^x$ and $s^z$ spin components.
The solutions for the system with the spin-orbit interaction generated by the electric field applied along the axis
of the nanotube/nanowire (case B) show that for the superposition of states with the same angular momenta
the $z$ spin component oscillates as a function of $z$ coordinate with the period that depends on the electric field.
We have also found that the longitudinal electric field generates the spin oscillation beats
localized along the nanotube/nanowire axis.
The position of these beats and consequently the maximal value of the average spin can be controlled by the bias voltage.

In summary, we have proposed the all-electrical mechanism of the electron spin manipulation based on the
spin-orbit interaction in semiconductor nanotubes and cylindrical nanowires.

\section*{Acknowledgments}
This work has been partly supported by the National Science Centre, Poland, under grant DEC-2011/03/B/ST3/00240.
P. W. has been supported by the Polish Ministry of Science and Higher Education and its grants for
Scientific Research.

\appendix
\section{Derivation of the effective 2D Hamiltonian}
\label{append}

We consider the cylindrical nanotube with the radius $r_0$ and the thickness in radial direction equal to $2\Delta r$
(in the case of cylindrical nanowire $2\Delta r$ is the thickness of charge accumulation layer).
The three-dimensional (3D) Hamiltonian for the single electron in the cylindrical nanotube/nanowire
with the spin-orbit interaction has the form
\begin{eqnarray}
\hat{H}_{3D}&=& \bigg [ -\frac{\hbar ^2}{2m} \left ( \frac{\partial ^2}{\partial r^2 }+ \frac{1}{r} \frac{\partial}{\partial r}
+\frac{1}{r^2} \frac{\partial }{\partial \varphi ^2} + \frac{\partial ^2}{\partial z^2} \right )\nonumber \\
&+& V(r,\varphi,z) \bigg ] \hat{\mathbb{I}} +\hat{H}^{3D}_{SO} \;,
\label{eq:a1}
\end{eqnarray}
where $r$, $\varphi$, and $z$ are the cylindrical coordinates,
$V(r,\varphi,z)$ is the potential energy,
and $\hat{H}^{3D}_{SO}=(\gamma / \hbar ) \vec{\sigma} \cdot (\nabla{V} \times \vec{p})$
 is the 3D spin-orbit interaction Hamiltonian.

If the potential energy possesses the rotational symmetry, i.e.,
$V(r,\varphi,z) = V(r,z)$, the spin-orbit interaction Hamiltonian can be expressed as
\begin{eqnarray}
&&\hat{H}^{3D}_{SOI}
=i \gamma
\bigg [
\left (
\begin{array}{cc}
0 & \frac{1}{r}\frac{\partial V}{\partial z} e^{-i\varphi} \frac{\partial}{\partial \varphi}\\
\frac{1}{r}\frac{\partial V}{\partial z} e^{i\varphi} \frac{\partial}{\partial \varphi} & 0
\end{array}
\right )
 \nonumber \\
&&+\left (
\begin{array}{cc}
0 & -i e^{-i \varphi } \left (\frac{\partial V}{\partial r} \frac{\partial}{\partial z}
- \frac{\partial V}{\partial z} \frac{\partial}{\partial r} \right )\\
i e^{i \varphi } \left (\frac{\partial V}{\partial r} \frac{\partial}{\partial z}
- \frac{\partial V}{\partial z} \frac{\partial}{\partial r} \right ) & 0
\end{array}
\right ) \nonumber \\
&& \left (
\begin{array}{cc}
- \frac{1}{r}\frac{\partial V}{\partial r}\frac{\partial}{\partial \varphi}  & 0\\
0 &  \frac{1}{r}\frac{\partial V}{\partial r}\frac{\partial}{\partial \varphi}
\end{array}
\right )
\bigg ] \;.
\end{eqnarray}

In the considered problem, the potential energy $V(r,z)$ is separable, i.e.,
$V(r,z)=V_r(r)+V_z(z)$, where $V_r(r)$ is responsible for confining the electron near the surface of the nanowire
and $V_z(r)$ is the potential energy of the electron in the axially directed electric field.
Assuming that the radial quantum states are spin-degenerate, the one-electron wave function can be
 take on in the form
\begin{equation}
  \Psi(r,\varphi,z)=\mathcal{R}(r) \psi(\varphi,z) \;,
  \label{Psi}
\end{equation}
where  $\mathcal{R}(r)$ is the radial wave function and $\psi(\varphi,z)$ is the spinor.\\
If the confinement of the electron in the nanotube is strong, we can assume
that the electron occupies the radial ground state $\mathcal{R}_{gs}(r)$.
The ground state with the definite parity with respect to $r=r_0$
has the following property:
\begin{equation}
\langle \mathcal{R}_{gs}| \frac{\partial}{\partial r} |\mathcal{R}_{gs} \rangle = 0 \;.
\end{equation}
Multiplying both the sides of Schr\"odinger equation with Hamiltonian (\ref{eq:a1}) by $\mathcal{R}^{\ast}_{gs}(r)$
and integrating over $r$ in range $[r_0-\Delta r, r_0+\Delta r]$ we obtain
\begin{eqnarray}
&&\bigg [ -\frac{\hbar ^2}{2m}
\bigg ( \bigg \langle \frac{1}{r^2} \bigg \rangle \frac{\partial }{\partial \varphi ^2} + \frac{\partial ^2}{\partial z^2} \bigg )
+E_0+ V_z(z) \bigg ] \hat{\mathbb{I}} \psi(\varphi,z) \nonumber \\
&& +\hat{H}_{SO}(r,\varphi,z)\psi(\varphi,z)=E\psi(\varphi,z) \;,
\end{eqnarray}
where
\begin{eqnarray}
&& \bigg \langle \frac{1}{r^2} \bigg \rangle =\langle\mathcal{R}_{gs}|\frac{1}{r^2}|\mathcal{R}_{gs} \rangle \;, \label{r0}\\
&&E_0=\langle\mathcal{R}_{gs}|-\frac{\hbar ^2}{2m} \left ( \frac{\partial ^2}{\partial r^2 }
+ \frac{1}{r} \frac{\partial}{\partial r} +V_R(r) \right )|\mathcal{R}_{gs}\rangle \;, \\
&&\hat{H}_{SO}=\langle\mathcal{R}_{gs}|\hat{H}_{3D,SO}|\mathcal{R}_{gs}\rangle \;.
\end{eqnarray}
For the strictly 2DEG   $\Delta r \rightarrow 0$, which allows us to perform the integration in (\ref{r0}).
For the normalized  $\mathcal{R}_{gs}(r)$, the integration in (\ref{r0}) is equal to $1/r_0^2$.
In the present paper, we treat energy $E_0$ as the reference energy and put it equal to 0.
Finally, we obtain the effective 2D Hamiltonian in the form
\begin{eqnarray}
\hat{H}=\left [ -\frac{\hbar ^2}{2m} \left ( \frac{1}{r_0^2} \frac{\partial ^2}{\partial \varphi ^2} +
\frac{\partial ^2}{\partial z ^2} \right ) +V_z(z) \right ] \hat{\mathbb{I}}  + \hat{H}_{SO},
\end{eqnarray}
where
\begin{eqnarray}
 \hat{H}_{SO}&=&
\left (
\begin{array}{cc}
0 & e \frac{\partial V}{\partial z} \frac{i\gamma}{r_0}e^{-i\varphi}\frac{\partial}{\partial \varphi} \\
e \frac{\partial V}{\partial z} \frac{i\gamma}{r_0} e^{i\varphi}\frac{\partial}{\partial \varphi}  & 0
\end{array}
\right )  \nonumber \\
&+&
\left (
\begin{array}{cc}
0 &  e^{-i\varphi} \gamma \frac{\partial V}{\partial r} \big |_{r=r_0}\frac{\partial}{\partial z}\\
- e^{i\varphi} \gamma \frac{\partial V}{\partial r}\big |_{r=r_0} \frac{\partial}{\partial z} & 0
\end{array}
\right ) \nonumber \\
&-&
\left (
\begin{array}{cc}
\frac{i\gamma}{r_0}\frac{\partial V}{\partial r}\big |_{r=r_0}\frac{\partial}{\partial \varphi} & 0\\
0 & -\frac{i\gamma}{r_0}\frac{\partial V}{\partial r}\big |_{r=r_0}\frac{\partial}{\partial \varphi}
\end{array}
\right ) \; .
\end{eqnarray}


\begin{thebibliography}{10}
\expandafter\ifx\csname url\endcsname\relax
  \def\url#1{\texttt{#1}}\fi
\expandafter\ifx\csname urlprefix\endcsname\relax\def\urlprefix{URL }\fi
\expandafter\ifx\csname href\endcsname\relax
  \def\href#1#2{#2} \def\path#1{#1}\fi

\bibitem{Datta}
S.~Datta, B.~Das, Appl. Phys. Lett. 56 (1990) 665.

\bibitem{Schliemann}
J.~Schliemann, J.~C. Egues, D.~Loss, Phys. Rev. Lett. 90 (2003) 146801.

\bibitem{Szumniak}
P.~Szumniak, S.~Bednarek, Phys. Rev. Lett. 109 (2012) 107201.

\bibitem{Berg}
J.~W.~G. van-den Berg, S.~Nadj-Perge, V.~S. Pribiag, S.~R. Plissard, E.~P.
  A.~M. Bakkers, S.~M. Frolov, L.~P. Kouwenhoven, Phys. Rev. Lett. 110 (2013)
  066806.

\bibitem{Dresselhause}
G.~Dresselhaus, Phys. Rev. 100 (1955) 580.

\bibitem{Rashba}
Y.~A. Bychkov, R.~E. I, Sov. Phys. JEPT 39 (1984) 78.

\bibitem{Nitta}
J.~Nitta, T.~Akasaki, H.~Takayanagi, T.~Enoki, Phys. Rev. Lett. 78 (1997) 1335.

\bibitem{Matsuyama}
T.~Matsuyama, R.~K{\"u}rsten, C.~Messer, U.~Merkt, Phys. Rev. B 61 (2000)
  15588.

\bibitem{Grundler}
D.~Grundler, Phys. Rev. Lett. 84 (2000) 6074.

\bibitem{Nazmitdinov}
R.~G. Nazmitdinov, K.~N. Pichugin, M.~Valin-Rodriguez, Phys. Rev. B 79 (2009)
  193303.

\bibitem{Nowack}
K.~C. Nowack, F.~H.~L. Koppens, Y.~V. Nazarov, L.~M.~K. Vandersypen, Science
  318 (2007) 1430.

\bibitem{Meier}
L.~Meier, G.~Salis, I.~Shorubalko, E.~Gini, S.~Scho{\"o}n, K.~Enslin, Nat.
  Phys. 3 (2007) 650.

\bibitem{NadjPerge}
S.~Nadj-Perge, V.~S. Pribiag, J.~W.~G. van~der Berg, K.~Zuo, S.~R. Plissard,
  E.~P. A.~M. Bakkers, S.~M. Frolov, L.~P. Kouwenhoven, Phys. Rev. Lett. 108
  (2012) 166801.

\bibitem{NadjPerge2}
S.~Nadj-Perge, S.~M. Frolov, J.~W.~W. van Tilburg, J.~Danon, V.~Nazarov, Yu,
  R.~Algra, P.~A.~M. Bakkers, E, L.~P. Kouwenhoven, Phys. Rev. B 81 (2010)
  201305(R).

\bibitem{Pfund}
A.~Pfund, I.~Shorubalko, R.~Leturcq, K.~Ensslin, Appl. Phys. Lett. 89 (2006)
  252106.

\bibitem{Ferrari2}
G.~Ferrari, G.~Cuoghi, Phys. Rev. Lett. 100 (2008) 230403.

\bibitem{Vorobov}
A.~B. Vorob'ev, K.~J. Friedland, H.~Kostial, R.~Hey, U.~Jahn, E.~Wiebicke,
  S.~Y. Yukecheva, V.~Y. Prinz, Phys. Rev. B 75 (2007) 205309.

\bibitem{Mendach}
S.~Mendach, O.~Schumacher, H.~Welsch, C.~Heyn, W.~Hansen, Appl. Phys. Lett. 88
  (2006) 212113.

\bibitem{Friedland}
K.~J. Friedland, R.~Hey, H.~Kostial, A.~Riedel, K.~H. Ploog, Phys. Rev. B 75
  (2007) 045347.

\bibitem{Lauhon}
L.~J. Lauhon, M.~S. Gudiksen, D.~Wang, C.~M. Lieber, Nature 420 (2002) 57.

\bibitem{Martensson}
T.~Martensson, M.~Borgst{\"o}m, W.~Seifert, B.~J. Ohlsson, L.~Samuelson,
  Nanotechnology 14 (2003) 1255.

\bibitem{Bjork}
M.~T. Bj{\"o}rk, B.~J. Ohlsson, T.~Sass, A.~I. Persson, C.~Thelander, M.~H.
  Megnusson, K.~Deppert, L.~R. Wallenberg, L.~Samuelson, Nano Lett 2 (2002) 87.

\bibitem{Ferrari}
G.~Ferrari, A.~Bertoni, G.~Goldoni, E.~Molinari, Phys. Rev. B 78 (2008) 115326.

\bibitem{Kleiner}
A.~Kleiner, Phys. Rev. B 67 (2003) 155311.

\bibitem{Wenk}
P.~Wenk, S.~Ketteman, Phys. Rev. B 81 (2010) 125309.

\bibitem{Wohlman}
O.~Entin-Wohlman, A.~Aharony, Y.~Tokura, Y.~Avishai, Phys. Rev. B 81 (2010)
  075439.

\bibitem{Bringer}
A.~Bringer, T.~Sch{\"a}pers, Phys. Rev. B 83 (2012) 115305.

\bibitem{Roulleau}
P.~Roulleau, T.~Choi, S.~Riedi, T.~Heinzel, I.~Shorubalko, T.~Ihn, K.~Ensslin,
  Phys. Rev. B 81 (2010) 155449.

\bibitem{Winkler}
R.~Winkler, Spin-orbit coupling effects in two-dimensional electron and hole
  systems, Springer Tracts in Modern Physics 191.

\end{thebibliography}
\end{document}